\documentclass[aps,pra,twocolumn]{revtex4}
\usepackage{longtable}
\usepackage{graphicx}
\usepackage{dcolumn}
\usepackage{amsmath}
\usepackage{amssymb}
\usepackage{subfigure}
\usepackage{float}
\usepackage{epstopdf}

\def\beq{\begin{equation}}
\def\eeq{\end{equation}}

\begin{document}
\title{ Effect of orbital-overlap dependence in density functionals}
\author{Jianwei Sun, Bing Xiao, and Adrienn Ruzsinszky}
\affiliation{Department of Physics and Quantum Theory Group, Tulane University, New Orleans, Louisiana 70118, USA}
\date{\today}

\begin{abstract}
 The semilocal meta generalized gradient approximation (MGGA) for the exchange-correlation functional of Kohn-Sham (KS) density functional theory can yield accurate ground-state energies simultaneously for atoms, molecules, surfaces, and solids, due to the inclusion of kinetic energy density as an input. We study for the first time the effect and importance of the dependence of MGGA on the kinetic energy density through the dimensionless inhomogeneity parameter, $\alpha$, that characterizes the extent of orbital overlap. This leads to a simple and wholly new MGGA exchange functional, which interpolates between the single-orbital regime, where $\alpha=0$, and the slowly varying density regime, where $\alpha \approx 1$, and then extrapolates to $\alpha \to \infty$. When combined with a variant of the Perdew-Burke-Erzerhof (PBE) GGA correlation, the resulting MGGA performs equally well for atoms, molecules, surfaces, and solids. 

\end{abstract}

\maketitle

 Kohn-Sham (KS) density functional theory (DFT) \cite{KS, Parr_Yang} is one of the most widely used methods in condensed matter physics and quantum chemistry. In this theory,
the exchange-correlation energy $E_{\rm xc}$ as a functional of the electron spin densities $n_{\uparrow}({\bf r})$ and $n_{\downarrow}({\bf r})$ must be approximated. Semilocal approximations (e.g., Refs.~\cite{PW92, SPS_PRB_2010, PBE, PBEsol, AM05, TPSS, PRCCS,ZT_JCP_2006}) of the form
\beq
E_{\rm xc}[n_{\uparrow}, n_{\downarrow}]=\int d^3r n \epsilon_{\rm xc}(n_{\uparrow},n_{\downarrow}, \nabla n_{\uparrow},\nabla n_{\downarrow},\tau_{\uparrow},\tau_{\downarrow})
\label{semi_local}
\eeq
require only a single integral over real space and so are practical even for large molecules or unit cells. In Eq.~\eqref{semi_local}, $\nabla n_{\uparrow,\downarrow}$ are the local gradients of the spin densities, $\tau_{\uparrow,\downarrow}=\sum_k \left |\nabla \psi_{k{\uparrow,\downarrow}} \right |^2/2$ the kinetic energy densities of the occupied KS orbitals $\psi_{k\sigma}$ of spin $\sigma$, and $ \epsilon_{\rm xc}$ the approximate exchange-correlation energy per electron. All equations are in atomic units.  Semilocal approximations can be reasonably accurate for the near-equilibrium and compressed ground-state properties of ‘‘ordinary’’ matter, where neither strong correlation nor long-range van der Waals interaction is important. They can also serve as a base for the computationally more-expensive fully nonlocal approximations needed to describe strongly correlated systems~\cite{PSTS_PRA_2008} and soft matter~\cite{DRSLL_PRL_2004}.

The meta generalized gradient approximation (MGGA) is the highest semilocal rung of the so-called Jacob's ladder in DFT \cite{Jacob_Ladder}. In addition to the spin densities $n_{\uparrow}$ and $n_{\downarrow}$ that are used in local spin density approximation (LSDA)~\cite{KS, PW92, SPS_PRB_2010} and their gradients that are further included in generalized gradient approximation (GGA)~\cite{PBE, PBEsol, AM05}, MGGA includes the kinetic energy density that can be used, as in the revised Tao-Perdew-Staroverov-Scuseria (revTPSS) MGGA~\cite{PRCCS}, to distinguish the single-orbital regions from the orbital-overlap regions.
However, the dependence of MGGAs on the kinetic energy density is understood much less than that on the density gradient, which MGGA inherits from GGA.  Such understanding is highly demanded by the construction of not only MGGAs theirselves but also fully nonlocal approximations that are based on MGGAs. Therefore, this could be largely beneficial to expediting the shift in DFT from the dominant GGAs to the generally more accurate and computationally comparable MGGAs~\cite{PRCCS, SMRKP}. In this article, we will show the importance and effect of the $\tau$-dependence, leading to a new MGGA that respects a tight Lieb-Oxford bound~\cite{LO_IJQC_1981, OC_JCP_2006} and performs equally well for atoms, molecules, surfaces, and solids.

\begin{figure}
\subfigure[]{\label{figure:Fx_alpha}\includegraphics[width=7cm]{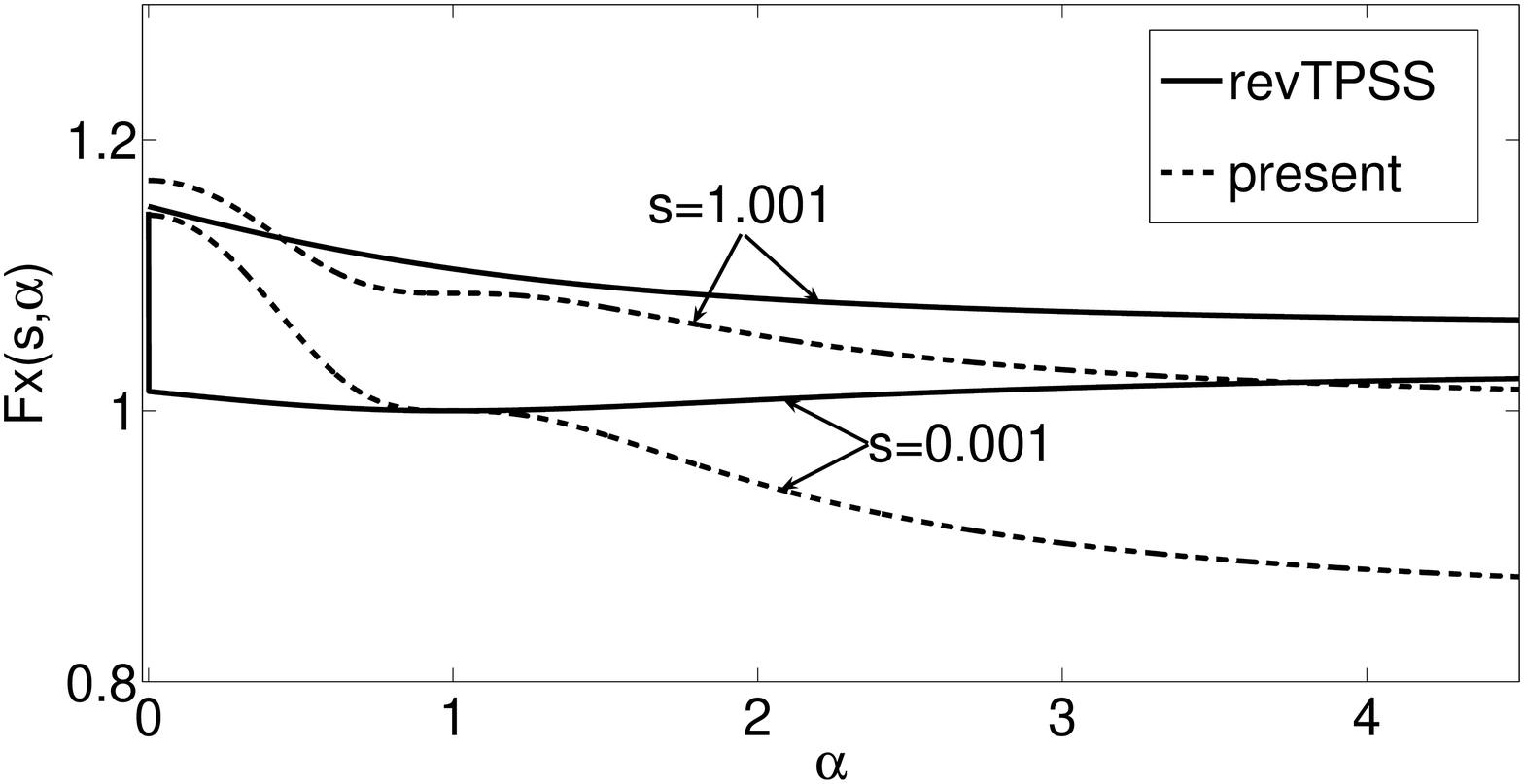}}
\hfill
\subfigure[]{\label{figure:Fx_s}\includegraphics[width=7cm]{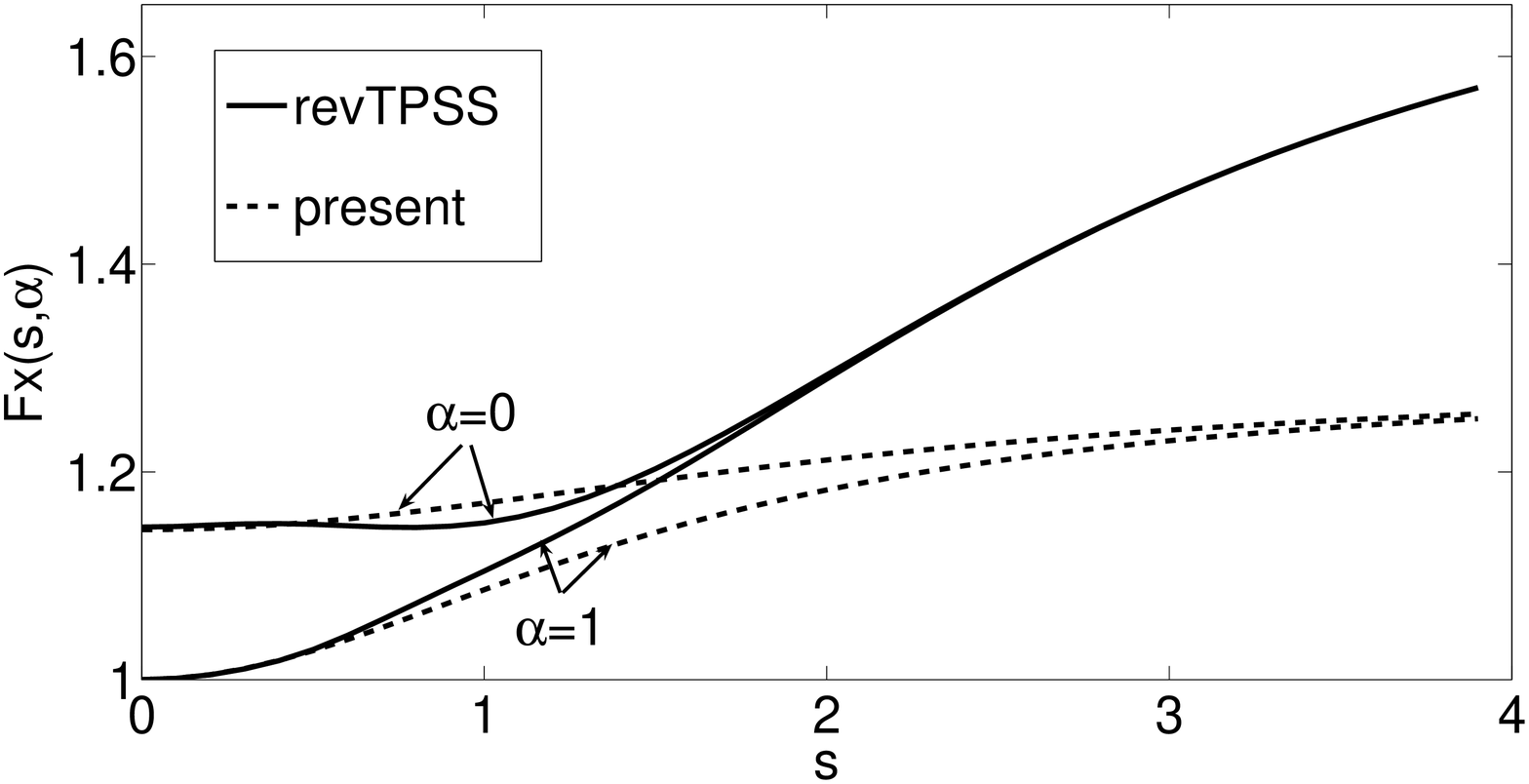}}

\caption{Exchange enhancement factors vs. $\alpha$ (a), and $s$ (b).}
\end{figure}

 The semilocal exchange energy of a spin-unpolarized density can be written as:
\beq
E_x^{\rm sl}[n]=\int d^3 r n \epsilon_x^{\rm unif}(n)F_x(p, \alpha).
\label{eq:Ex_sl}
\eeq
Here, $\epsilon_x^{\rm unif}(n)=-\frac{3}{4\pi}(\frac{9\pi}{4})^{1/3} / r_s$ is the exchange energy per particle of a uniform electron gas with $r_s=(4 \pi n /3)^{-1/3}$, $p=|\nabla n|^2/[4(3 \pi ^2)^{2/3} n ^{8/3}]=s^2$, and $\alpha=(\tau-\tau^W)/\tau^{\rm unif} \geqslant 0$. In the latter, $\tau=\sum_\sigma \tau_\sigma$, $\tau^W=\frac{1}{8}|\nabla n|^2/n$ is the  von Weizs$\ddot{{\rm a}}$cker kinetic energy density, and $\tau^{\rm unif}=\frac{3}{10}(3 \pi^2)^{2/3}n^{5/3}$ is the orbital kinetic energy density of the uniform electron gas. The expression for the spin-polarized case follows from the spin-scaling relation~ \cite{PKZB}. The enhancement factors $F_x$ (which equals 1 in LSDA) of GGAs, independent of $\alpha$, are {\it often} made monotonically increasing with $s$ as in the standard PBE GGA~\cite{PBE}, or for a large range of $s$~\cite{PW91,LG_PRA_1992,VMT_JCP_2009}, and therefore favor less compact systems than LSDA does (e.g., lowering the energy of a collection of dissociated atoms relative to that of the molecules, or lowering a surface energy, or enlarging lattice constants of solids). The revTPSS MGGA includes the extra ingredient, $\alpha$, and recovers the exact exchange energy of the ground state density of the hydrogen atom and the finiteness of the exchange potential at nuclei, where $\alpha=0$ (single-orbital regime). Then, at $\alpha \approx 1$ (slowly-varying density regime), it restores the second order gradient expansion for a wide range of density and further recovers the fourth order gradient expansion of a slowly varying density.  Therefore, we believe that the revTPSS enhancement factor is accurate for small $s$ around $\alpha \approx 1$. However, in the construction of revTPSS, there is no other constraint to guide the functional approaching from $\alpha=0$ to $\alpha \approx 1$. revTPSS also has an order-of-limits problem \cite{PTSS_JCP_2004, regTPSS}---the enhancement factor has different values when different sequences of the limits $s \to 0$ and $\alpha \to 0$ are taken, as shown in Fig. \ref{figure:Fx_alpha}. 


 Here, we propose a simple exchange enhancement factor that disentangles $\alpha$ and $p$ by a means of separability assumption,
\beq
F_x^{\rm int}(p, \alpha)=F_x^{\rm 1}(p)+f(\alpha)[F_x^{\rm 0}(p)-F_x^{\rm 1}(p)],
\label{eq:Fx_MGGA}
\eeq
where $F_x^{\rm 1}(p)=F_x^{\rm int}(p, \alpha=1)=1+\kappa-\kappa/(1+\mu^{\rm GE}p/\kappa)$ and $F_x^{\rm 0}(p)=F_x^{\rm int}(p, \alpha=0)=1+\kappa-\kappa/(1+(\mu^{\rm GE}p+c)/\kappa)$. $F_x^{\rm int}(p, \alpha)$ interpolates between $F_x^{\rm 0}(p)$ and $F_x^{\rm 1}(p)$ through $f(\alpha)=(1-\alpha^2)^3/(1+\alpha^3+\alpha^6)$, which is chosen to guarantee for the functional the second order gradient expansion, good exchange jellium surface energies, and the Hartree-Fock exchange energy of  the 12-electron hydrogenic density ($1s^22s^22p^63s^2$) with the nuclear charge $Z=1$~\cite{SSPTD_PRA_2004} ($Z=1$ in the following discussions if not mentioned otherwise). See Ref.~\onlinecite{SUPP} for the derivatives of $F_x^{\rm int}(p, \alpha)$ with respect to $p$ and $\alpha$ which are needed for the selfconsistent implementation.  For a slowly varying density  where $\alpha \approx 1$, the second term of the left hand side of Eq.~\ref{eq:Fx_MGGA} is negligible and of order $\nabla^6$. $F_x^{\rm int}(p, \alpha)$ then reduces to $F_x^{\rm 1}(p)$ and  recovers the second order gradient expansion with the first principle coefficient $\mu^{\rm GE}=10/81$~\cite{AK_PRB_1985} as in PBEsol~\cite{PBEsol}.
$c=0.28771$ and $\kappa=0.29$ are two parameters fixed by the exchange energies of the hydrogen atom {\it where $\alpha=0$} and the 12-electron hydrogenic density which is used to guide the functional from $\alpha=0$ to $\alpha \approx 1$. They also deliver excellent exchange energies for other hydrogenic densities (see Ref.~\onlinecite{SUPP}). No point in the (c, $\kappa$) parameter space could be found without violating the loose Lieb-Oxford bound~\cite{LO_IJQC_1981} ({\it i.e.,} $\kappa \leqslant \kappa^{LO}=0.804$~\cite{PBE}) if we use a spin-unpolarized hydrogenic density with electron number less than 12. 
The obtained $\kappa=0.29$ suggests a tight Lieb-Oxford bound and leads to a very flat $F_x^{\rm 0}(p)$ as shown in Fig.~\ref{figure:Fx_s}. The resulting small derivative at nuclei, $\frac{d F_x^{\rm int}(p, \alpha=0)}{d s} |_{s=0.376}=0.22$, comes close to satisfying the finiteness constraint on the exchange potential at nuclei. Compared to revTPSS, this much simpler form doesn't have the order-of-limits problem while recovering regions of small $s$ around $\alpha \approx 1$ of revTPSS as shown in Fig.~\ref{figure:Fx_alpha}. 
In the following discussion, we combine this exchange functional with the variant of the PBE correlation (denoted as vPBEc), where $\beta(r_s)=0.066725(1+0.1 r_s)/(1+0.1778 r_s)$ as used in revTPSS~\cite{PRCCS}. Although the vPBEc is not one-electron self-correlation free, its error is small (about 0.006 Ha for the H atom), which is usually largely cancelled out of atomization energies involving H. It has an accurate correlation energy for a two-electron ion of nuclear charge $Z \to \infty$, which is -0.0479 ~\cite{PBE}, better than -0.0510 from TPSS and -0.0527 from revTPSS in comparison with the exact value -0.0467~\cite{SSPTD_PRA_2004}. And it also has accurate correlation jellium surface energies~\cite{PRCCS}. The resulting MGGA respects a tight Lieb-Oxford bound with $F_{\rm xc} \leqslant 2.137$, while the loose Lieb-Oxford bound is 2.273~\cite{LO_IJQC_1981, OC_JCP_2006}.

 From the experience of GGAs~\cite{PBE, PBEsol}, we know the faster an enhancement factor grows with $s$, the more preference of the functional towards less compact systems. Since the enhancement factor of the present exchange functional as shown in Fig.~\ref{figure:Fx_s} is largely depressed for a large range of $s$ compared to that of revTPSS, it then seems to be a reasonable guess from the experience that the present exchange functional with the vPBEc would give too small lattice constants for solids and too high atomization energies for molecules. However, our results (given later) show a different scenario that our MGGA performs equally well for atoms, molecules, surfaces, and solids. The seeming contradiction between the preformance and Fig.~\ref{figure:Fx_s} is well resolved by resorting to the $\alpha$-dependence shown in Fig.~\ref{figure:Fx_alpha}, which is much less understood. In previous constructions and analysis of MGGAs~\cite{TPSS, PRCCS, SMRKP}, it is emphasized that $\alpha$ enables MGGAs to distinguish the single-orbital regions from the orbital-overlap regions. Here, we further stress that the monotonically decreasing dependence of an enhancement factor on $\alpha$ is qualitatively equivalent to the monotonically increasing $s$-dependence, which explains the seeming contradiction and will be rationalized by the following two observations.

\begin{figure}
\includegraphics[width=7cm]{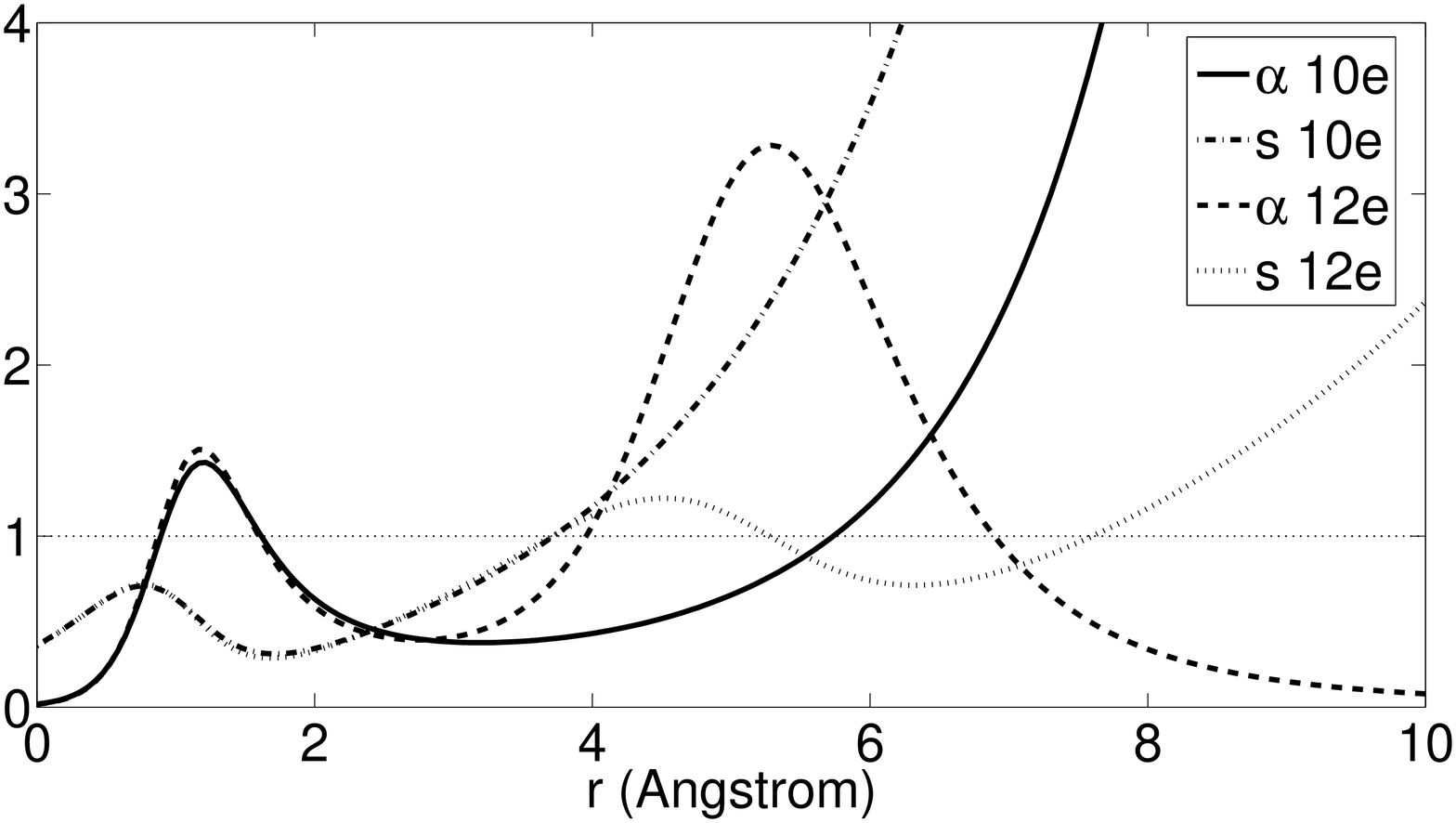}
\caption{The s and $\alpha$ distributions for the 10- and 12-electron hydrogenic densities. The line y=1 is plotted to help roughly identify the shell ($\alpha<1$) and intershell ($\alpha>1$) regions.}
\label{figure:s_alpha_10e_12e}
\end{figure}

  The first is on the changes in $s$ and $\alpha$ distributions from the 10-electron hydrogenic density ($1s^22s^22p^6$) to the 12-electron one ($1s^22s^22p^63s^2$)~\cite{SSPTD_PRA_2004}, and on the correlation between the changes. Note these two densities are spherically symmetric. Fig.~\ref{figure:s_alpha_10e_12e} shows the $s$ and $\alpha$ distributions of these two densities. The shell structure can be easily recognized and roughly identified by $\alpha$ with $\alpha < 1$ for shell regions and $\alpha > 1$ for intershell regions. We can tolerate the confusion caused by this definition for the tail regions, where $\alpha$ could be 0 or $\gg$ 1 as shown in  Fig.~\ref{figure:s_alpha_10e_12e}, since the tail regions are energetically less important. 
When an electron is in the intershell region, and $\alpha$ is large even where $s$ is small, the electron's exchange hole is probably not centered close to the electron, but is spread out over the smaller inner shell and the larger outer shell. This spreading of the hole will make the exchange energy density in the intershell region less negative than it would be for a slowly-varying or uniform density. One can imagine $F_x<1$, as can happen in the present exchange functional for small $s$ and large $\alpha$ that is shown in Fig.~\ref{figure:Fx_alpha}.

When two $3s$ hydrogenic electrons are added to the 10-electron hydrogenic density, part of the outermost shell region ($\alpha < 1$) changes into the intershell region ($\alpha > 1$), which is associated with a decrease of $s$. Although it's not certain that an increase of $\alpha$ is always associated with a decrease of $s$ during the formation of the intershell of an atom, it's very likely for the intershell region between the outermost core and the valence of an atom within a solid, an important region for determining the lattice constants of solids~\cite{HTBSL_PRB_2009, FBPS_PRB_1998} (as will be discussed in the second observation). Then, the presumable correlation between $s$ and $\alpha$ in the intershell regions of a solid suggests that monotonically decreasing $\alpha$-dependence of an enhancement factor has the qualitatively same effect as monotonically increasing $s$-dependence does for these regions. 


\begin{center}
\noindent
\begin{table}[htb]
\caption{Error statistics of lattice constants ($\AA$) of the SL20 solids \cite{SMCRHKKP_PRB_2011}
and atomization energies (kcal/mol) of the AE6 set \cite{LT_JPCA_2003}. 
See the text for the definitions of $F_x^{\rm 0 int}$, $F_x^{\rm \star int}$, and $F_x^{\rm int}$. }
\centering
\vspace{2mm}
\begin{tabular}{l c c c c c c}
\hline\hline

	&	&LSDA&PBEsol&$F_x^{\rm 0 int} $	&$F_x^{\rm \star int}$ & $F_x^{\rm int}$\\
\hline						
	&ME	&-0.081&-0.012	&-0.079	&-0.126	&0.016\\
SL20	&MAE	&0.081&0.036	&0.079	&0.126	&0.023\\
\hline
	&ME	&77.4&35.9	&55.1	&22.0	&0.6\\
AE6	&MAE	&77.4&35.9	&55.1	&22.6	&5.5\\
\hline
\end{tabular}

\label{table:falpha}
\end{table}
\end{center}

The second observation concerns the variations of the lattice constants of the set of 20 solids (SL20) \cite{SMCRHKKP_PRB_2011} and the atomization energies of the AE6 molecule set \cite{LT_JPCA_2003} in response to changes of the $\alpha$ dependence of the enhancement factor. In order to show the dramatic effect of the $\alpha$ dependence on the structural and thermochemichal properties and to deduce its origin, $f(\alpha)$ is compared to two variants. The first variant is $f^0(\alpha)=0$, which is independent of $\alpha$ and actually results in the PBEsol GGA but with different $\kappa$ and $\beta(r_s)$. The second one is $f^\star(\alpha)=f(\alpha)*h(\alpha)$ with $h(\alpha)=2/[e^{(\alpha-1)/\gamma}+1]-1$. $h(\alpha)$ is equal to 1 for $\alpha < 1$ and to -1 for $\alpha > 1$, respectively, if $\gamma \to 0$. Here, we choose $\gamma=0.1$ for numerical reasons.  Compared to $f(\alpha)$, $f^\star(\alpha)$ flips the $\alpha$ dependence for $\alpha>1$ from monotonically decreasing to monotonically increasing, and therefore favors more the regions with $\alpha>1$. The choice of the demarcation point at $\alpha=1$ is natural in view of the first observation and because it helps to satisfy the second order gradient expansion. $f(\alpha)$ and the two variants---whose curves as functions of $\alpha$ are given in Ref.~\onlinecite{SUPP}---result in three different enhancement factors, $F_x^{\rm int}$, $F_x^{\rm 0 int}$, and $F_x^{\rm \star int}$. 


\begin{center}
\noindent
\begin{table}[htb]
\caption{Error statistics for various density functionals. The LSDA and PBE enthalpies of formation are from Ref.~\onlinecite{PRCCS}. The lattice constants for M06L are from Ref.~\onlinecite{ZT_JCP_2008}, where 18 solids, mostly overlapping with the SL20 set, were tested. 
a: Ref.~\onlinecite{ZT_JCP_2008} }
\centering
\vspace{2mm}
\begin{tabular}{p{0.51 in} p{0.51 in} p{0.51 in} p{0.51 in} p{0.51 in} p{0.51 in} }
\hline\hline
	&LSDA	&PBE	&M06L	&revTPSS	&present  \\ [0.5ex]
\hline
\multicolumn{6}{c}{Exchange energies (Ha) of rare gas atoms}\\ 
ME	&2.274	&0.219	&0.204	&0.291	&0.068\\ 
MAE	&2.274	&0.219		&0.210	&0.293	&0.111\\ 
\multicolumn{6}{c}{Atomization energies (kcal/mol) of the AE6 molecules}\\ 
ME	&77.4	&12.4	&3.2	&3.3	&0.6\\ 
MAE	&77.4	&15.5	&4.2	&5.9	&5.5\\ 
\multicolumn{6}{c}{Dissociation energies (kcal/mol) of the W6 water clusters}\\ 
ME	&5.2&0.0	&-0.2	&-1.0	&0.0\\ 
MAE	&5.2&0.3	&0.4	&1.0    	&0.1\\ 
\multicolumn{6}{c}{Enthalpies of formation (kcal/mol) of the G3 molecules}\\ 
ME	&-121.9	&-21.7	&-1.6	&-3.6	&-1.6\\ 
MAE	&121.9	&22.2		&5.2	&4.8	       &8.3\\ 
\multicolumn{6}{c}{Jellium surface exchange energies ($\%$)}\\ 
MRE	&45.8	&-20.9	&-75.9	&-1.0	&-7.3\\ 
MARE	&45.8	&20.9	&75.9	&2.2	&8.0\\ 
\multicolumn{6}{c}{Jellium surface exchange-correlation energies ($\%$)}\\ 
MRE	&-0.4	&-3.1	&24.5	&2.6&-0.3\\ 
MARE	&0.4	&3.1	&24.5	&2.6	&1.6\\ 
\multicolumn{6}{c}{Lattice constants ($\AA$) of the SL20 solids}\\ 
ME	&-0.081	&0.051		&$0.015^a$&0.015	&0.016\\ 
MAE	&0.081	&0.059	&0.071$^a$	&0.033&0.023\\ 

\hline							
\hline
\end{tabular}

\label{table:results}
\end{table}
\end{center}

Table~\ref{table:falpha} shows the mean error (ME) and the mean absolute error (MAE) of the lattice constants of the SL20 set \cite{SMCRHKKP_PRB_2011}, and the atomization energies of the AE6 set \cite{LT_JPCA_2003} from LSDA, PBEsol, and variants of $F_x^{\rm int}$. The alleviation, from LSDA to $F_x^{\rm 0 int}$ and then to $F_x^{\rm int}$, of the overestimation in the atomization energies and of the underestimation in the lattice constants, suggests that the built-in monotonically increasing $s$-dependence and monotonically decreasing $\alpha$-dependence in the enhancement factors reduce the preference of LSDA towards compact systems. 
However, from $F_x^{\rm int}$ to $F_x^{\rm \star int}$, where only the monotonicity of the $\alpha$-dependence in the range of [1, $\infty$] is flipped, the solids in the SL20 set are drastically shrunk to such a surprising degree that the lattice constants of $F_x^{\rm \star int}$ are significantly smaller than even those of LSDA. Since the important region in terms of determining the lattice constant of a solid for a functional has been identified \cite{HTBSL_PRB_2009, FBPS_PRB_1998} to be the intershell region of the constituent atoms between the outermost core and the valence regions, the drastic shrinkage from $F_x^{\rm int}$ to $F_x^{\rm \star int}$ is a strong indication that the shell and intershell regions are associated with $\alpha < 1$ and $\alpha > 1$, respectively. Compressing a solid turns part of the outermost core and the valence regions ($\alpha < 1$) into intershell regions ($\alpha > 1$) between them, which $F_x^{\rm \star int}$ favors more than $F_x^{\rm int}$ does. 
The absence of the monotonically decreasing $\alpha$-dependence in $F_x^{\rm 0 int}$, leading to the shrinkage of the solids, could be compensated by enhancing the monotonically increasing dependence on $s$, as PBEsol does by using $\kappa^{LO}$. This implies a decrease of $s$ during the formation of the intershell regions and thus corroborates the {\it correlation} between $s$ and $\alpha$ during the formation of the intershell region observed in Fig.~\ref{figure:s_alpha_10e_12e}. Therefore, in $F_x^{\rm int}$, both the monotonically increasing $s$-dependence and the monotonically decreasing $\alpha$-dependence have the effect of penalizing the formation of the intershell regions and enlarging the lattice constants.
Similar deterioration is also found for the atomization energies of the AE6 set for $F_x^{\rm \star int}$ compared to $F_x^{\rm int}$. Unlike for solids, $F_x^{\rm \star int}$ still significantly improves the atomization energies of the AE6 set over $F_x^{\rm 0 int}$, and thus LSDA, suggesting that the $\alpha$-dependence in the range of [0, 1] has stronger influence in atoms and molecules than in solids. Remarkably, the monotonically decreasing $\alpha$-dependence used in $F_x^{\rm int}$ significantly improves the overestimated atomization energies of the AE6 set of $F_x^{\rm 0 int}$ to an excellent accuracy level with MAE of 5.4 kcal/mol. This implies that monotonically decreasing $\alpha$-dependence is in general able to make a functional favor less compact systems, as does monotonically increasing $s$-dependence.

Now, let's turn to the results for atoms, molecules, surfaces, and solids, which are summarized briefly in Table~\ref{table:results} in terms of the ME and MAE, or their relative analogs MRE and MARE. See Ref.~\onlinecite{SUPP} for full details of Table~\ref{table:results}. Table~\ref{table:results} shows that the use of the exact exchange energy of the 12-electron hydrogenic density guarantees excellent exchange energies for atoms, resulting in good atomization energies for this simple functional. In all categories shown in Table~\ref{table:results}, the present functional outperforms the standard PBE GGA (PBE is slightly better than the present functional for the cohesive energies of the 20 SL20 solids as shown in Ref.~\onlinecite{SUPP}). Within the MGGA level, The heavily parameterized M06L \cite{ZT_JCP_2006} predicts excellent atomization energies and dissociation energies for the W6 set, at which it aims during the construction. However, it is significantly wrong for the jellium surface energies contributed from the exchange and correlation terms, separately or together. The too-large M06L jellium surface exchange-correlation energies
imply that metal bulks are overstabilized, consistent with the too-small lattice constants of main group simple metals, e.g., Na and Al ~\cite{ZT_JCP_2008}. The present functional and revTPSS are more balanced for different categories and therefore more robust.
Compared to revTPSS, the present functional is better for the SL20 solids, comparable for the jellium surface exchange-correlation energies but worse for the exchange part alone, and worse for the G3 molecules. 
However, the present functional predicts the most accurate dissociation energies of the W6 water clusters among the functionals, implying a good description for the hydrogen bond. 


In summary, we have for the first time studied the effect and importance of the dependence of computationally-efficient semilocal MGGAs on the kinetic energy density through the dimensionless inhomogeneity parameter $\alpha$, and presented a new MGGA exchange functional that disentangles $\alpha$ from the reduced density gradient $s$ by the means of separability assumption. By varying the $\alpha-$dependence in the exchange functional, we showed that the formation of the intershell region between the outermost core and the valence of an atom within a solid is associated with an increase of $\alpha$ and a decrease of $s$, suggesting that monotonically decreasing $\alpha$-dependence of an enhancement factor is qualitatively equivalent to monotonically increasing $s$-dependence for these intershell regions. This has a significant impact on the construction of MGGAs and the MGGA-based nonlocal approximations, as exemplified by the present MGGA---which is overall comparable in performance, but quite different and much simpler in form, compared to the sophisticated revTPSS MGGA, and thus demonstrates the flexibility and the rich structure of MGGA brought by the extra ingredient of the kinetic energy density.

{\bf Acknowledgments} JS thanks John P. Perdew, G$\acute{\rm a}$bor I. Csonka, and Stephen E. Glindmeyer for helpful discussions. JS and BX are supported by NSF under Grant No. DMR08-54769. AR acknowledges support from the NSF under NSF Cooperative Agreement No. EPS-1003897. Portions of this research were conducted with high performance computational resources provided by the Louisiana Optical Network Initiative (http://www.loni.org/).

\end{document}